\documentclass[10pt, conference]{IEEEtran}
\IEEEoverridecommandlockouts
\usepackage{textcomp}
\usepackage{xcolor}
\def\BibTeX{{\rm B\kern-.05em{\sc i\kern-.025em b}\kern-.08em T\kern-.1667em\lower.7ex\hbox{E}\kern-.125emX}}
\usepackage{cite}
\usepackage{amsmath,amssymb,amsfonts}
\usepackage[english]{babel}
\usepackage{amsthm}
\usepackage{graphicx}
\usepackage[caption=false,font=footnotesize]{subfig}
\usepackage{algpseudocode}
\usepackage{algorithm} 
\newtheorem{theorem}{Theorem}

\newtheorem{defn}{Definition}

\newtheorem{lem}{Lemma}
\usepackage{stfloats}
\usepackage{blindtext}
\usepackage{mathrsfs}
\usepackage[compact]{titlesec}
\begin{document}

\title{\huge Bayesian Game Formulation of Power Allocation in Multiple Access Wiretap Channel with Incomplete CSI
}
\author{
\IEEEauthorblockN{Basharat Rashid\IEEEauthorrefmark{1},~Majed Haddad \IEEEauthorrefmark{2}, and Shahid M Shah\IEEEauthorrefmark{1}}
\IEEEauthorblockA{\IEEEauthorrefmark{1}Dept. of ECE, National Institute of Technology Srinagar, J\&K, India\
Email: rashidbasharat@yahoo.com, shahidshah@nitsri.ac.in}
\IEEEauthorblockA{\IEEEauthorrefmark{2}CERI/LIA, University of Avignon, Avignon, France\
Email: majed.haddad@univ-avignon.fr}
}
\maketitle
\begin{abstract}
In this paper, we address the problem of distributed power allocation in a $K$ user fading multiple access wiretap channel, where global channel state information is limited, i.e., each user has knowledge of their own channel state with respect to Bob and Eve but only knows the distribution of other users' channel states. We model this problem as a Bayesian game, where each user is assumed to selfishly maximize his average \emph{secrecy capacity} with partial channel state information. In this work, we first prove that there is a unique Bayesian equilibrium in the proposed game. Additionally, the price of anarchy is calculated to measure the efficiency of the equilibrium solution. We also propose a fast convergent iterative algorithm for power allocation. Finally, the results are validated using simulation results.
\end{abstract}


\section{Introduction}
Fading Multiple Access Channel (F-MAC) is well studied in wireless communication as it models the uplink scenario in a cellular system, where $K$ users are trying to communicate with a base station. Further, optimal resource allocation techniques in an F-MAC are extensively established in the literature. Wyner and Shamai first studied F-MAC, assuming that only the receiver has access to channel state information (CSI) \cite{ref_article1}. This work was later extended in \cite{ref_article2} to include the assumption that transmitters have global CSI, with the authors utilizing the polymatroidal structure of the capacity region to derive power control policies. In \cite{ref_article3}, the problem of resource allocation was framed as a complete information game, with transmitters having global CSI. Power allocation in F-MAC was first approached as an incomplete information game in \cite{ref_article4}, with the assumption of individual CSI and global knowledge of the channel state distribution. The same setup was extended to a two-tier cellular network in \cite{Nguyen2015Stackelberg}.

 In the literature, the concept of F-MAC has been extended to the fading multiple access wiretap channel (F-MAC-WT), where an eavesdropper attempts to intercept information in addition to a legitimate receiver. In \cite{tekin2007secrecy}, F-MAC-WT with global CSI at the transmitters is studied. In \cite{shah2012achievable}, the authors proposed a power control scheme for F-MAC-WT without any knowledge of the channel state of the eavesdropper. In \cite{shah2016resource} the authors assumed local channel state information, where each user is assumed to know his channel state with respect to Bob and Eve, but the transmitter relies on low-rate feedback from the receiver. In \cite{Zhifan2021friendly}, authors employed a zero-sum game to maximize the \emph{secrecy rate} of the network.

In this study, we employ Bayesian game theory to explore the optimal distributed power allocation in a single input single output (SISO) F-MAC-WT with limited global CSI. Specifically, we consider a $K$ user uplink scenario, where each user possesses knowledge of his channel state in relation to Bob and Eve but only knows the distribution of other users channel states. The Bayesian game proved to be a useful tool to investigate this incomplete information scenario, where each rational user aims to maximize their achievable \emph{secrecy rate} with restricted knowledge of fading channel gains. Further, we use the concept of diagonal concavity \cite{ref_article9} to demonstrate the existence of a unique Bayesian equilibrium (BE) in this context and propose an iterative algorithm to characterize it. Finally, the price of anarchy (POA) is calculated as a metric to measure the efficiency of the equilibrium state.

As cellular technologies continue to advance, networks are becoming denser (for example 5G/6G communication systems), resulting in an increased demand for high data rates and decentralized cellular systems \cite{tinh2022practical}. In this regard, the proposed Bayesian game theoretical framework aligns with the concept of secure ``self-organizing networks", which reduces operational complexity and eliminates the need for global CSI.

The rest of the paper is organized as follows: In Section II, we introduce the channel model, and in Section III, we develop the Bayesian game setup and prove the uniqueness of BE. Subsequently, in Section IV, we characterize the BE and evaluate its efficiency. Finally, in Section V, we provide simulation results to establish our theoretical claims.

\section{Fading MAC-WT Model}
We consider a Gaussian wiretap channel for an uplink scenario where $K$ users are attempting to communicate with a legitimate receiver (Bob) while being intercepted by an adversary (Eve). The signal transmitted by the $k^{th}$ user is represented as $X_{k}$, while the channel gain of the $k^{th}$ user in relation to Bob and Eve is represented by $\widetilde{H}_{k}$ and $\widetilde{G}_{k}$ respectively. The received signals by Bob and Eve are represented by $Y$ and $Z$, respectively, and are given as:
\vspace{-0.2cm}
\begin{align}
Y=\sum_{k=1}^{K}\widetilde{H}_{k}X_k+N_{k}^{b}~,~~
 Z=\sum_{k=1}^{K}\widetilde{G}_kX_k+N_{k}^{e},
 \label{eq:1}
\end{align}
where $N_{k}^{b}$ and $N_{k}^{e}$ are zero-mean additive white Gaussian noise with variance $\sigma^{2}$. It is assumed that the noise $N_{k}^{b}$ is independent of channel gain $\widetilde{H}_k$, also the channel realization $\widetilde{H}_k$ is independent of $\widetilde{H}_l, ~ \forall l \neq k$. Similar assumptions hold for the channel gain $\widetilde{G}_k$ and the noise $N_{k}^{e}$. Additionally, $\widetilde{H}_k$ is assumed to be independent of $\widetilde{G}_k,~\forall k $.
In this paper, we examine the transmission of wireless signals in a fast-fading environment, where the channel's coherence time is relatively small compared to the time delay. It is further assumed that the channel is stationary and ergodic during the transmission. For a degraded Gaussian wiretap channel where the signal received by Bob is stronger than Eve, the achievable rate in this scenario for secure communication is characterized by \emph{secrecy capacity}\cite{Leung1978Gaussian} and for the assumed fast fading setup we use the notion of \emph{ergodic secrecy capacity}, which is defined as:
\vspace{-0.1cm}
\begin{align}
C^{s}_k=\mathbb{E}_{\textbf{H},\textbf{G}}\Bigg[\Bigg(C_k^{b}-C_k^{e}\Bigg)^+\Bigg],
\label{eq:2}
 \end{align} 
where $\textbf{H}=\{H_{1},\dots,H_{K}\}$, $\textbf{G}=\{G_{1},\dots,G_{K}\}$ and the notation is defined as $(x)^+=\max(0,x)$. Further, $C^{b}_k$ and $C^{e}_k$ are instantaneous \emph{Shannon capacity} for Bob and Eve, respectively, and for non-successive decoding at the receiver, they are given by
\vspace{-0.3cm}
\begin{align}
 C_k^b=\log_{2}\Bigg(1+\frac{\mathit{H}_{k}P_{k}}{\sigma^2+\sum\limits_{l\neq k,l=1}^{K}\mathit{H}_{l}P_{l}}\Bigg)
 \label{eq:3}\\
 C_k^e=\log_{2}\Bigg(1+\frac{\mathit{G}_{k}P_{k}}{\sigma^2+\sum\limits_{l\neq k,l=1}^{K}\mathit{G}_{l}P_{l}}\Bigg),
 \label{eq:4}
\end{align}
where $H_k=\widetilde{H}_k^{2}$, $G_k=\widetilde{G}_k^{2}$ and $P_k$ is the transmitted power.
\section{Bayesian game setup for F-MAC-WT}
Consider an incomplete information scenario, where each transmitter can perfectly estimate his channel gain $H_k$ and $G_k$ with Bob and Eve, respectively, but does not know the value of the channel state of other users, i.e., $\textbf{H}_{-k}=\{H_{1},\dots,H_{k-1},H_{k+1},\dots,H_{K}\}$ and $\textbf{G}_{-k}=\{G_{1},\dots,G_{k-1},G_{k+1},\dots,G_{K}\}$ are unknown at $k^{th}$ transmitter. However, the distribution of each $H_l$ and $G_{l}$, $\forall l\neq k$ is known at the $k^{th}$ transmitter. In such a scenario, each user aims to maximize his \emph{ergodic secrecy capacity} subject to the average power constraint $\overline{P}_k$. Hence, the following optimization problem is sought for each user
 \begin{align}
 \max_{P_k}~ \mathbb{E}_{\textbf{H},\textbf{G}}\Bigg[\Bigg(C_k^b-C_k^e\Bigg)^+ \Bigg ],~~~~~~~~~~~~\label{eq:5}\\ 
\text{\it{s.t.}}~\mathbb{E}_{H_k,G_k}[P_k(H_k,G_k)]\leq \overline{P}_k,~P_k(H_k,G_k)\geq 0 \nonumber 
\end{align} 
For the given setup we define
\begin{align}
C_k^b=\log_{2}\Bigg(1+\frac{\mathit{H}_{k}P_{k}(H_k,G_k)}{\sigma^2+\sum\limits_{l\neq k,l=1}^{K}\mathit{H}_{l}P_{l}(H_l,G_l)}\Bigg)
\label{eq:6}\\
C_k^e=\log_{2}\Bigg(1+\frac{\mathit{G}_{k}P_{k}(H_k,G_k)}{\sigma^2+\sum\limits_{l\neq k,l=1}^{K}\mathit{G}_{l}P_{l}(H_l,G_l)}\Bigg),
\label{eq:7}
\end{align}
where $P_k(H_k,G_k)$ is the transmitted power, which is the function of the user's own channel gains with Bob and Eve. However, the optimization problem (\ref{eq:5}) is dependent on the power strategies of all the transmitting users, which is not common knowledge. Thus, in order to achieve optimal power allocation, each user needs to adapt his power output depending on his guess of the strategies of all other users. Bayesian game theory is a well-suited framework to address the problem of incomplete information, as it provides a structured approach for users to modify their strategies based on their assumptions about other users.

Consider each user's channel gains are i.i.d, discrete with $L$ states i.e, $H_{k}=\{h_{1},\dots,h_{L}\}$ and $G_{k}=\{g_{1},\dots,g_{L}\}$, where $h_i$ and $g_j$, $i,j \in [1,L]$ are the gains over a coherence time. Therefore, the $K-$ user Bayesian game for F-MAC-WT can be completely characterized as: 
 \begin{align}
 \mathcal{G}_{F-MAC-WT}\triangleq \left(\mathcal{K},\mathcal{T},\mathcal{P},\mathcal{Q},\mathcal{U}\right),
 \label{eq:8}
 \end{align}
 where each element of a tuple is defined as:
 \begin{itemize}
 \item Player set: $\mathcal{K}$ is the set of users, $\mathcal{K}=\{1,\dots,K\}$.
 \item Type set: $\mathcal{T}=\mathcal{T}_{1}\times \dots \times \mathcal{T}_{K}$ (`$\times$' defines Cartesian product), where $\mathcal{T}_{k}= H_k \times G_k $ is the set of tuples. A player's type is defined by its channel gains, that is $(h_i, g_j)\in \mathcal{T}_{k}$.
 \item Action set: $\mathcal{P}=\mathcal{P}_{1}\times \dots \times \mathcal{P}_{K}$, where $\mathcal{P}_k=[0,P_k^{max}]$. A player's action is defined as his transmit power, that is, $P_k\in \mathcal{P}_k$.
 \item Probability set: $\mathcal{Q}=\mathcal{Q}_{1}\times\dots\times\mathcal{Q}_{K} $, where $\mathcal{Q}_k= A_k \times B_k$ with $A_k=\{\alpha_1,\dots,\alpha_L\}$ and $B_k=\{\beta_1,\dots,\beta_L\}$, where $\alpha_i\triangleq Pr\{H_k=h_i\}$ and $\beta_j\triangleq Pr\{G_k=g_j\}$.
 \item Utility function set: $\mathcal{U}=\{U_1,\ldots,U_K\}$, where $U_k$ is $k^{th}$ user's achievable rate.
 \begin{align}
 \hspace*{-1.1cm}
 U_k(P_1,\ldots,P_K)=\Big(C^b_k-C^e_k\Big)^{+}
 \label{eq:9}
 \end{align}
 \end{itemize}
\begin{defn}[Bayesian equilibrium] The strategy profile $P^{*}(\cdot )=\{P_{k}^{*}(\cdot )\}_{k\in \mathcal{K}}$ is a pure strategy Bayesian equilibrium, if for all $k\in \mathcal{K}$, and for all $P_{k}(\cdot)\in \mathcal{P}_{k}$ and $\mathbf{P}_{-k}(\cdot)\in \mathcal{P}_{-k}$, we have:
 \begin{align}
 \overline{U}_k(P^*_k,\mathbf{P}_{-k}^*)\geq \overline{U}_k(P_k,\mathbf{P}_{-k}^*),
 \label{eq:10}
 \end{align}
 where $\overline{U}_k=\mathbb{E}_{\textbf{H},\textbf{G}}[U_k]$ is the average utility.
\end{defn}
\subsection{Existence of unique Bayesian equilibrium }
\begin{theorem}
 For a degraded fading multiple access wiretap channel, there exists a unique Bayesian equilibrium for $K$ user game $\mathcal{G}_{F-MAC-WT}$.
\end{theorem}
\emph{Proof.}
 Since the strategy space $P_k$ is convex, compact, and non-empty for each user $k$, the utility function $U_k$ is continuous with respect to both $P_k$ and $\textbf{P}_{-k}$. Additionally $U_k$ is concave in $P_k$ for any $\textbf{P}_{-k}$\cite{ref_article8}.
 
 Based on the above conditions, according to \cite{ref_article9}: a non-cooperative game has unique equilibrium if the non-negative weighted sum of the utility functions is diagonally strictly concave. Before we proceed further, we will state the following definition.

 \begin{defn} [Diagonally strictly concave] To ensure that a non-negative sum function $f(\textbf{X},\lambda)=\sum_{i=1}^N \lambda_i \phi_i(\textbf{X})$ is diagonally strictly concave for a fixed vector $\lambda \in \mathbb{R}^{N\times 1}_{+}$ and any vector $\textbf{X}:=\prod\limits_{i=1}^{N} X_{i}\in \mathbb{R}^{\sum m_{i}}$, where $X_i \in \mathbb{R}^{m_{i}}$, a sufficient condition is that the symmetric matrix $[J(\textbf{X},\lambda)+J^{T}(\textbf{X},\lambda)]$ is negative definite, 
Where $J(\textbf{X},\lambda)$ is the Jacobian of 
 $\triangledown(\textbf{X},\lambda)$ with respect to $\textbf{X}$ and $J^{T}$ is transpose of $J$. $\triangledown(\textbf{X},\lambda)$ is called the pseudo gradient of $f(\textbf{X},\lambda)$ and is defined as:
 \begin{align}
 \triangledown(\mathbf{X},\lambda)\triangleq
 \begin{bmatrix}
 \lambda_1\triangledown_{1}\phi_1(\textbf{X})\\
 \vdots\\
 \lambda_N\triangledown_{N}\phi_N(\textbf{X})
 \end{bmatrix}
 \label{eq:11},
 \end{align}
 where $\triangledown_{i}\phi(\textbf{X})$ is the gradient of $\phi(\textbf{X})$ with respect to $X_i$.
 \end{defn}
 \begin{lem} The weighted non-negative sum of the average utilities $\overline{U}_k$ is diagonally strictly concave for $\lambda=c^+\mathbf{1}$, where $c^+$ is a positive scalar, $\mathbf{1}$ is a vector whose every entry is 1.
\end{lem}
\emph{Proof.} For the sake of clarity, we examine a two-user scenario, \textit{i.e.}, $K=2$. We emphasize that this is not a restriction of the proposed model. Let us first define $P_{k}\in \mathbb{R}^{M}$ as the strategy vector of $k^{th}$ user and $\textbf{P}:=\prod\limits_{k=1}^{2} P_{k}\in \mathbb{R}^{2M}$ as the action profile for the game. Furthermore, it is considered that $P_{ki},i=1,\dots,M$, is the $i^{th}$ strategy of the $k^{th}$ user, which corresponds to the value of channel states $H_{k}$ and $G_{k}$. For the assumed setup, we can define the average utility as follows:
\vspace{-0.5cm}
\begin{align}
\overline{U}_k(\mathbf{P})=&\sum_{j}\sum_{i}{\gamma_{j}}\gamma_{i}\Bigg[\log\Bigg(1+ \frac{H_{k}P_{ki}(H_{k},G_{k})}{\sigma^{2}+H_{l}P_{lj}(H_{l},G_{l})}\Bigg)\nonumber \\ &-\log\Bigg(1+\frac{G_{k}P_{ki}(H_{k},G_{k})}{\sigma^{2}+G_{l}P_{lj}(H_{l},G_{l})}\Bigg)\Bigg]^{+},~k\neq l
\label{eq:12}
\end{align} 
where $\gamma_{i}$ is the joint probability of different combinations of channel gains $H_k$ and $G_k$ and is related to the value of $\alpha_{i}$ and $\beta_{i}$. Similarly, $\gamma_{j}$ is defined for $H_l$ and $G_l$. Further, we will define the weighted, non-negative sum of the average payoffs as:
\vspace{-0.6cm}
\begin{align}
 f^u(\mathbf{P},\lambda)\triangleq \sum_{k=1}^{2}\lambda_k\overline{U}_k(\mathbf{P})
 \label{eq:13}
 \end{align}
Similar to (\ref{eq:11}), we define the pseudo gradient as:
\begin{align}
\triangledown^{U}
(\mathbf{P},\lambda)=
\begin{bmatrix}
c^{+}\frac{\partial \overline{U}_{1}(\textbf{P})}{\partial P_{11}} \\
\vdots\\
c^{+}\frac{\partial \overline{U}_{1}(\textbf{P})}{\partial P_{1M}} \\[6pt]
c^{+}\frac{\partial \overline{U}_{2}(\textbf{P})}{\partial P_{21}}\\
\vdots\\
c^{+}\frac{\partial \overline{U}_{2}(\textbf{P})}{\partial P_{2M}}
\end{bmatrix}
\label{eq:14}
\end{align}In order to evaluate each element of the column vector (\ref{eq:14}), we have $i^{th}$ utility for the $k^{th}$ user given as:
\vspace{-0.3cm}
 \begin{align}
 U(P_{ki})=&\Bigg[\log\Bigg(1+\zeta^{b}_{lj} P_{ki}(H_{k},G_{k})\Bigg)\nonumber \\
&-\log\Bigg(1+\zeta^{e}_{lj}P_{ki}(H_{k},G_{k})\Bigg)\Bigg]^{+},
\label{eq:15}
 \end{align}
where,
\vspace{-0.3cm}
\begin{align}
\zeta^{b}_{lj}=\frac{H_{k}}{\sigma^{2}+H_{l}P_{lj}(H_{l},G_{l})} , ~~\zeta^{e}_{lj}=\frac{G_{k}}{\sigma^{2}+G_{l}P_{lj}(H_{l},G_{l})} ~~
 \label{eq:16}
 \end{align}
Throughout, we will assume that the channel is degraded, i.e., $\zeta^{b}_{lj}>\zeta^{e}_{lj}$, this inequality captures the fact that the signal received by Bob is better than Eve. Otherwise, the achievable \emph{secrecy capacity} is zero. Therefore, the partial derivative of the average utility function, whenever it exists, is given by
 \begin{align}
 \frac{\partial \overline{U}_{k}(\textbf{P})}{\partial P_{ki}}=&\sum_{j}\gamma_{j}\gamma_{i}\Bigg(\frac{\zeta^{b}_{lj}}{1+\zeta^{b}_{lj}P_{ki}(H_{k},G_{k})}\nonumber\\&-\frac{\zeta^{e}_{lj}}{1+\zeta^{e}_{lj}P_{ki}(H_{k},G_{k})}\Bigg)
 \label{eq:17}
 \end{align}
Therefore the pseudo gradient can be written as:
 \begin{align}
& \triangledown^{U}(\mathbf{P},\lambda)= \nonumber \\
& \begin{bmatrix}
c^{+}\sum\limits_{j}\gamma_{j}\gamma_{1}\Bigg(\frac{\zeta^{b}_{2j}}{1+\zeta^{b}_{2j}P_{11}(H_{1},G_{1})}-\frac{\zeta^{e}_{2j}}{1+\zeta^{e}_{2j}P_{11}(H_{1},G_{1})}\Bigg)\\
 \vdots\\
c^{+}\sum\limits_{j}\gamma_{j}\gamma_{M}\Bigg(\frac{\zeta^{b}_{2j}}{1+\zeta^{b}_{2j}P_{1M}(H_{1},G_{1})}-\frac{\zeta^{e}_{2j}}{1+\zeta^{e}_{2l}P_{1M}(H_{1},G_{1})}\Bigg)\\
c^{+}\sum\limits_{i}\gamma_{1}\gamma_{i}\Bigg(\frac{\zeta^{b}_{1i}}{1+\zeta^{b}_{1i}P_{21}(H_{2},G_{2})}-\frac{\zeta^{e}_{1i}}{1+\zeta^{e}_{1i}P_{21}(H_{2},G_{2})}\Bigg)\\
 \vdots\\
c^{+}\sum\limits_{i}\gamma_{M}\gamma_{i}\Bigg(\frac{\zeta^{b}_{1i}}{1+\zeta^{b}_{1i}P_{2M}(H_{2},G_{2})}-\frac{\zeta^{e}_{1i}}{1+\zeta^{e}_{1i}P_{2M}(H_{2},G_{2})}\Bigg) \nonumber
 \end{bmatrix}
\end{align}
The second order partial derivatives of $\overline{U}_{k}(\textbf{P})$ are given in (\ref{eq:18})-(\ref{eq:20}).
\begin{figure*}[b]
\rule{\linewidth}{0.4pt}
\begin{align} 
\frac {\partial \overline{U}_{k}^{2}(\textbf{P})}{\partial P_{ki}^{2}}=c^{+} \sum_{j}\gamma_{j}\gamma_{i}\Bigg[\Bigg(\frac{\zeta^{e}_{lj}}{1+\zeta^{e}_{lj}P_{ki}(H_{k},G_{k})}\Bigg)^2-\Bigg(\frac{\zeta^{b}_{lj}}{1+\zeta^{b}_{lj}P_{ki}(H_{k},G_{k})}\Bigg)^2\Bigg],\forall i
\label{eq:18}
\end{align}
\begin{align} 
\frac {\partial \overline{U}_{k}^{2}(\textbf{P})}{\partial P_{lj}P_{ki}}=c^{+}\gamma_{j}\gamma_{i}\Bigg(\frac{\zeta^{e}_{lj}\zeta^{e}_{ki}}{(1+\zeta^{e}_{lj}P_{ki}(H_{k},G_{k}))(1+\zeta^{e}_{ki}P_{lj}(H_{l},G_{l}))}-\frac{\zeta^{b}_{lj}\zeta^{b}_{ki}}{(1+\zeta^{b}_{lj}P_{ki}(H_{k},G_{k})(1+\zeta^{b}_{ki}P_{lj}(H_{l},G_{l})}\Bigg), \forall i,j
\label{eq:19}
\end{align}
\begin{align}
\frac {\partial \overline{U}_{k}^{2}(\textbf{P})}{\partial P_{ki^{'}}\partial P_{ki}}= 0, ~~ \forall i^{'} \neq i
\label{eq:20}
\end{align}
\end{figure*}
Now, the Jacobian of $\triangledown^{U}(\mathbf{P},\lambda)$ with respect to $\textbf{P}$ is defined as:
\vspace{-0.3cm}
\begin{align}
 & J(\textbf{P},\lambda)= \nonumber \\
 &\begin{bmatrix}
 \frac {\partial \overline{U}_{1}^{2}(\textbf{P})}{\partial P_{11}^{2}} & \dots & \frac {\partial \overline{U}_{1}^{2}(\textbf{P})}{\partial P_{1M}\partial P_{11}} & \frac {\partial \overline{U}_{1}^{2}(\textbf{P})}{\partial P_{21}\partial P_{11}} &\dots & \frac {\partial \overline{U}_{1}^{2}(\textbf{P})}{\partial P_{2M}\partial P_{11}}\\
 \vdots & \ddots & \ddots & \ddots & \ddots &\vdots\\ 
 \frac {\partial \overline{U}_{2}^{2}(\textbf{P})}{\partial P_{11}\partial P_{2M}} & \dots & \frac {\partial \overline{U}_{2}^{2}(\textbf{P})}{\partial P_{1M}\partial P_{2M}} & \frac {\partial \overline{2}_{1}^{2}(\textbf{P})}{\partial P_{21}\partial P_{2M}} &\dots & \frac {\partial \overline{U}_{2}^{2}(\textbf{P})}{\partial P_{2M}^{2}}
 \end{bmatrix}
 \nonumber
\end{align}
 From (\ref{eq:16}) its evident that $\zeta_{lj}^b$ and $\zeta_{lj}^e$ are positive, now for degraded channel condition, i.e., $\zeta_{lj}^b>\zeta_{lj}^e$ from (\ref{eq:18}) we have:
 \vspace{-0.5cm}
\begin{align}
&\Bigg(\frac{\zeta^{e}_{lj}}{1+\zeta^{e}_{lj}P_{ki}(H_{k},G_{k})}\Bigg)^2-\Bigg(\frac{\zeta^{b}_{lj}}{1+\zeta^{b}_{lj}P_{ki}(H_{k},G_{k})}\Bigg)^2
 \nonumber\\
&=\frac{(\zeta^{e}_{lj})^2-(\zeta^{b}_{lj})^2+\zeta^{b}_{lj}\zeta^{e}_{lj}
P_{ki}(H_{k},G_{k})(\zeta^{e}_{lj}-\zeta^{b}_{lj})}{(1+\zeta^{e}_{lj}P_{ki}(H_{k},G_{k}))^2(1+\zeta^{b}_{ki}P_{lj}(H_{l},G_{l}))^2}<0
\end{align}
Similarly, we can prove (\ref{eq:19}) is also negative. Therefore, all the non-zero terms in $J(\textbf{P},\lambda)$ are negative, which proves that the symmetric matrix is negative definite, i.e., $\textbf{P}^{T}[J(\textbf{P},\lambda)+J^{T}(\textbf{P},\lambda)]\textbf{P}<0$.
 
Therefore, from definition (2), $f^u(\mathbf{P},\lambda)$ is diagonally strictly concave, which proves lemma (1). As we have proved $f^u(\mathbf{P},\lambda)$ is diagonally strictly concave, it follows from \cite{ref_article9} that there exists a unique Bayesian equilibrium for the game $\mathcal{G}_{F-MAC-WT}$. Although we have proven our results for a two-user scenario, it is evident that the same can be extended to any $K$.
\section{Characterization of Bayesian Equilibrium}
For the two-user setup of the proposed game $\mathcal{G}_{F-MAC-WT}$, let $P_{ij}^k=P^{k}(h_i,g_j)$ be the power allocated by the $k^{th}$ user corresponding to the specific channel gain $h_i$ and $g_j$. Considering that each user observes the same set of $L$ channel states with Eve and Bob, the power allocation vector for $k^{th}$ user is $\textbf{P}^k=\{P_{11}^{k},\dots P_{1L}^{k},\dots, P_{L1}^{k}, \dots, P_{LL}^{k} \}$ with the average power constraint $\overline{P}^k$. Further, for the degraded channel condition, the average utility of each player can be specified as:
\vspace{-0.3cm}
\begin{align}
&\overline{U}_k(\textbf{P}^k,\textbf{P}^l)=\nonumber\\ &\sum_{m=1}^{L}\sum_{n=1}^{{L}}\sum_{j=1}^{L}\sum_{i=1}^{L}\beta_{m}\alpha_{n}\beta_{j}\alpha_{i}\Bigg[\log\Bigg( 1+\frac{h_{i}P_{ij}^k}{\sigma^{2}+h_{n}P_{nm}^l}\Bigg)\nonumber\\
& \quad \quad \quad -\log\Bigg( 1+\frac{g_{j}P_{ij}^k}{\sigma^{2}+g_{m}P_{nm}^l}\Bigg)\Bigg],~l\neq k
\label{eq:22}
\end{align}
To find the best-response strategy of the $k^{th}$ player given the strategy profile $\textbf{P}^l$ of another player, we need to solve the following maximization problem
\vspace{-0.2cm}
\begin{align}
 \max_{\textbf{P}^k}~ \overline{U}_k(\textbf{P}^k,\textbf{P}^l),~~~~~~~ \label{eq:23}\\
 \text{\it{s.t.}}~\mathbb{E}_{H_k,G_k}[P_{ij}^k]\leq \overline{P}^k,~~P_{ij}^k\geq 0 \nonumber
 \end{align}
In this problem, the objective function $\overline{U}_k$ is a concave function in $\textbf{P}^k$, and the set of constraints is convex. Hence, the optimization problem can be classified as convex optimization, and the corresponding Lagrangian $\mathcal{L}$ is given by
\vspace{-0.3cm}
\begin{align*}
&\mathcal{L}=\\
\vspace{-0.3cm}
&\overline{U}_k(\textbf{P}^k,\textbf{P}^l)+\lambda^k \Bigg(\sum_{i=1}^{L}\sum_{j=1}^{L}\beta_{j}\alpha_{i}P_{ij}^k-\overline{P}^k\Bigg)-\sum_{i=1}^{L}\sum_{j=1}^{L}\emph{v}_{ij}^k P_{ij}^k 
\end{align*}
where $\lambda^{k}>0$ and $\nu_{ij}^{k}>0, \forall~ k,i,j$ are the Lagrange multipliers associated with inequality constraints. Therefore, the Karush-Kuhn-Tucker (KKT) conditions for optimization are necessary and sufficient for optimality and are given in (\ref{eq:24}) - (\ref{eq:25}).
\begin{figure*}[b]
\rule{\linewidth}{0.4pt}
\begin{align} 
\sum_{m=1}^{L}\sum_{n=1}^{{L}}\beta_{m}\alpha_{n}\beta_{j}\alpha_{i}\Bigg( \frac{h_{i}}{\sigma^{2}+h_{n}P_{nm}^l+h_{i}P_{ij}^k}-\frac{g_{i}}{\sigma^{2}+g_{n}P_{nm}^l+g_{i}P_{ij}^k}\Bigg)+\lambda^k\beta_{j}\alpha_{i}-\nu_{ij}^k=0, ~\forall i,j
\label{eq:24}
\end{align}
\begin{align}
\lambda^{(k)} \Bigg(\sum_{i=1}^{L} \sum_{j=1}^{L}\beta_{j}\alpha_{i}P_{ij}^k-\overline{P}^k\Bigg) = 0, ~~~~~\&~~~\nu_{ij}^k P_{ij}^k = 0, ~~\forall i,j
\label{eq:25}
\end{align}
\end{figure*}
For the degraded channel from (\ref{eq:24}), it is easy to verify that $\lambda^k>0$. Therefore, from (\ref{eq:25}) we have
\vspace{-0.3cm}
\begin{align}
 \sum_{i=1}^{L} \sum_{j=1}^{L}\beta_{j}\alpha_{i}P_{ij}^k=\overline{P}^k
 \label{eq:26}
 \vspace{-0.2cm}
\end{align}The above equation signifies that at Bayesian equilibrium, each user distributes his power across all the channel realizations depending on their distributions. Based on equation (\ref{eq:24}), the optimal solution for each user is dependent on the transmission power of other user's, which is not universally known. Therefore, in order to achieve the optimal power allocation, each user must individually optimize his transmission power based on his estimate of the power allocation of the other user. In equation (\ref{eq:22}), we define $\psi_{nm}^{b,k}=\sigma^{2}+h_{n}P_{nm}^l$ and $\psi_{nm}^{e,k}=\sigma^{2}+g_{n}P_{nm}^l$ as the aggregate interference observed by $k^{th}$ user in relation to Bob and Eve, respectively. Hence, for distributed power allocation by $k^{th}$ user, we propose an Algorithm (\ref{alg:loop}).
\vspace{-0.2cm}
\begin{algorithm}[H] 
\caption{Distributed Iterative Power Allocation}
\label{alg:loop}
\begin{algorithmic}[1]
\Statex
\State
\textbf{Input:} Set of channel states $H_k=\{h_1,\dots,h_L\}$ and $G_k=\{g_1,\dots,g_L\}$
\State Initialize $t=0$ , $ P_{ij}^{k(0)}=0 
~~\forall~k$ and $\forall ~ i,j$
\State \textbf{Repeat:} Until converge
\State $ t\gets t+1$
\For{$k \gets 1$ to $2$} 
 \For{$n \gets 1$ to $L$} 
 \For{$m \gets 1$ to $L$} 
 \State $\psi_{nm}^{b,k(t)}=\sigma^{2}+h_{n}P_{nm}^{l(t)}$ 
 \State $\psi_{nm}^{e,k(t)}=\sigma^{2}+g_{n}P_{nm}^{l(t)}$
 \EndFor
 \EndFor
 \vspace{-0.2cm}
\State \begin{align}
 [P_{11}^{k},\dots ,P_{1L}^{k},\dots, P_{L1}^{k}, \dots, P_{LL}^{k}] ~~~~~~~~~~~~~~~~~~~~ \nonumber \\ ~~~~=\underset{\mathbf{E}_{H_k,G_k}[P_{ij}^k]\leq \overline{P}^k,~P_{ij}^k\geq 0}{argmax} ~\overline{U}_k(\textbf{P}^k,\textbf{P}^l) \nonumber
\end{align}
\EndFor
\end{algorithmic}
\end{algorithm}
\vspace{-0.2cm}
This algorithm assumes that the players will play the same game multiple times, without any memory of past games or knowledge of future events. In every round, each player will choose their own optimal strategy by solving KKT conditions which are determined by the game's current state.
Hence, if each player's power allocation $P_{ij}^k$ is determined by applying the single-player KKT conditions while treating other player's signals as noise, the set $\textbf{P}^k$ is guaranteed to be a BE of the original game $\mathcal{G}$. Therefore, we can infer that the Algorithm (\ref{alg:loop}) converges through iteration to a BE point.
\subsection{Efficiency of Bayesian equilibrium}
In order to evaluate the system's efficiency when operating at the equilibrium state we calculate PoA as a metric. We first define social welfare as the sum of individual \emph{ergodic secrecy} rates
\vspace{-0.5cm}
\begin{align}
U_{\emph{sum}}(\textbf{P}) =\sum_{k=1}^{2}\overline{U}_k(\textbf{P}),
\label{eq:27}
\end{align}
where $\textbf{P}:=\prod\limits_{k=1}^{2} \textbf{P}^{k}$ and $\overline{U}_k(\cdot)$ is given in (\ref{eq:22}). We define $R_{\emph{sum}}^{BE}$ as the sum rate at the BE, where each user optimizes his rate in a decentralized manner by solving (\ref{eq:23}) based on the estimate of other user's strategies. Consider $C_{\emph{sum}}^{OPT}$ as the social optimal that can be achieved with centralized control and feedback signaling between the base station and the users. 
\begin{defn} [Price of Anarchy] The Price of Anarchy (PoA) is the ratio between the network sum rate at unique Bayesian equilibrium and the social optimal.
 \begin{align}
 PoA=\frac{R_{\emph{sum}}^{BE}}{C_{\emph{sum}}^{OPT}}
 \label{eq:28}
 \end{align} 
\end{defn}
\noindent The problem of maximizing the overall network rate in a centralized approach can be expressed as:
\vspace{-0.2cm}
\begin{align}
&~~~~~~~~\max_{\textbf{P}}~ {U}_{\emph{sum}}(\textbf{P}),\label{eq:29}\\
 \text{\it{s.t.}}&~\mathbb{E}_{H_k,G_k}[P_{ij}^k]\leq \overline{P}^k~\forall~ k, ~ P_{ij}^k\geq 0\nonumber 
 \end{align}
\noindent Solving the above optimization problem is challenging because the objective function is non-convex with respect to $\textbf{P}$. However, we can address this problem by utilizing the lower bound given in \cite{ref_article10}, namely
\begin{align}
 \omega \log(z)+ \Omega \leq \log(1+z),
 \label{eq:30}
\end{align}
where the bound is said to be tight for a chosen $z_{o}$ if
\begin{align}
 \omega =\frac{z_{o}}{1+z_{o}},~~~~ \Omega = \log(1+z_{o})-\frac{z_{o}}{1+z_{o}}\log(z_{o})
 \label{eq:31}
\end{align}
Applying relaxation (\ref{eq:30}) to (\ref{eq:22}) with the transformation of variable $\Tilde{P}_{ij}^{k}=\log(P_{ij}^{k})$, we can convert the optimization problem (\ref{eq:29}) into a Difference of Convex (DC) problem, namely 
\vspace{-0.6cm}
\begin{align}
 &~~~~~~~\min_{\Tilde {\textbf{P}}}~ \sum_{k=1}^{2} \mathcal{F}_{k}(\Tilde {\textbf{P}})- \mathcal{H}_{k}(\Tilde {\textbf{P}}),~~~ \label{eq:32} \\
&\text{\it{s.t.}}~\mathbb{E}_{H_k,G_k}[2^{\Tilde{P}_{ij}^k}]\leq \overline{P}^{k}~~\forall~ k ,~~\Tilde{P}_{ij}^{k}\geq 0  \nonumber
 \end{align}
 where,
 \vspace{-0.4cm}
 \begin{align}
\mathcal{F}_{k}(\Tilde {\textbf{P}})=&-\sum_{m=1}^{L}\sum_{n=1}^{{L}}\sum_{j=1}^{L}\sum_{i=1}^{L}\beta_{m}\alpha_{n}\beta_{j}\alpha_{i} \Bigg\{\omega_{ij,nm}^{b,k}\Bigg(\log(h_{i})\nonumber\\
&+\Tilde{P}_{ij}^{k}-\log(\sigma^{2}+h_{n}2^{\Tilde{P}_{nm}^l})\Bigg) +\Omega_{ij,nm}^{b,k}\Bigg\}
\label{eq:33}\\
 \mathcal{H}_{k}(\Tilde {\textbf{P}})=&-\sum_{m=1}^{L}\sum_{n=1}^{{L}}\sum_{j=1}^{L}\sum_{i=1}^{L}\beta_{m}\alpha_{n}\beta_{j}\alpha_{i}\Bigg\{\omega_{ij,nm}^{e,k} \Bigg(\log(g_{i})\nonumber\\
 &+\Tilde{P}_{ij}^{k}
-\log(\sigma^{2}+g_{n}2^{\Tilde{P}_{nm}^l})\Bigg)+ \Omega_{ij,nm}^{e,k}\Bigg\}
\label{eq:34}
\end{align}
The above two equations are clearly convex since the log-sum-exp function is convex. Hence, we can solve (\ref{eq:32}) using the generic DCA algorithm \cite{ref_article11}. However, as we are optimizing the lower bound, we can use the iterative procedure given in \cite{ref_article10} to improve the bound, and finally do the reverse transformation $P_{ij}^{k}=2^{\Tilde {P}_{ij}^{k}}$ to get the value of actual power.
\section{Simulations}
\subsection{Convergence of Ergodic rate}
\vspace{-0.38cm}
\begin{figure}[H]
\centering
\begin{minipage}[t]{0.49\linewidth}
\centering
\includegraphics[width=\linewidth]{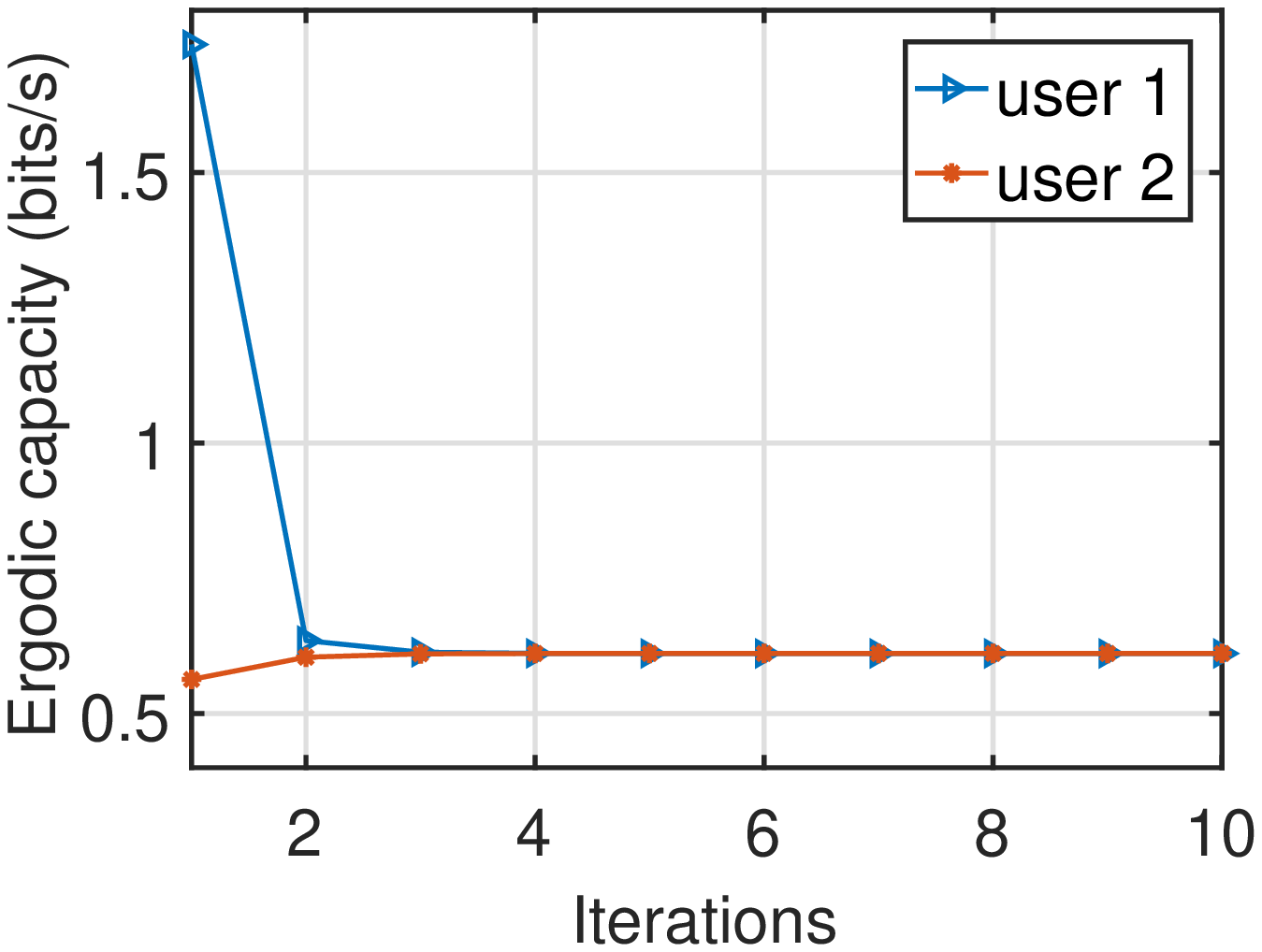}
\\{\footnotesize (a)}
\end{minipage}
\hfill
\begin{minipage}[t]{0.49\linewidth}
\centering
\includegraphics[width=\linewidth]{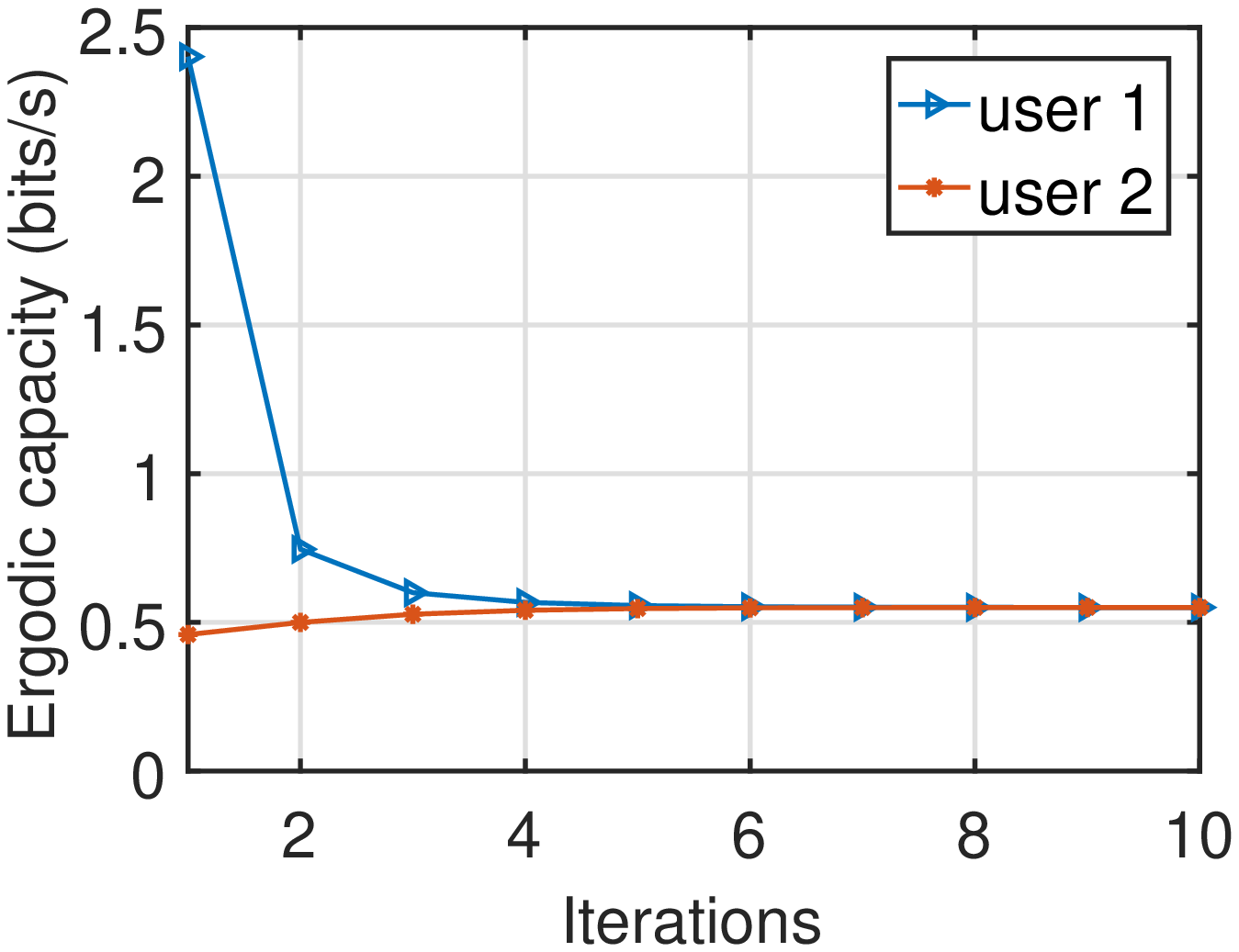}
\\{\footnotesize (b)}
\end{minipage}
\caption{Convergence of Algorithm (\ref{alg:loop}) for different SNR values:\\
(a)  $SNR=1$.
(b) $SNR=5$.}
\end{figure}
\vspace{-0.38cm}
To ensure a clear understanding, we consider a two-user scenario with two-channel states concerning Eve and Bob i.e., $L=2$, with system parameters $\alpha_{i}=0.5$ and $\beta_{j}=0.5$,$\forall i,j$ also $H_k=\{2, 3.5\}, G_k=\{0.2, 0.3\}$ and $\sigma^{2}=1$ for both users. For an SNR of 1 and 5, figures (1a) and (1b) above show the rapid convergence of \emph{ergodic secrecy capacity} to a unique solution by using the proposed Algorithm (\ref{alg:loop}). Here, SNR is defined as the ratio of average power $(\Bar{P}^k)$ and noise power $(\sigma^{2})$. Numerical simulations confirm the convergence of the Algorithm (\ref{alg:loop}). However, it is observed that the convergence rate slows down at higher SNR levels.
\subsection{Ergodic sum-rate efficiency}
\begin{figure*}[htbp]
\vspace{-0.2cm}
\centering
\begin{minipage}[t]{0.32\linewidth}
\centering
\includegraphics[width=\linewidth]{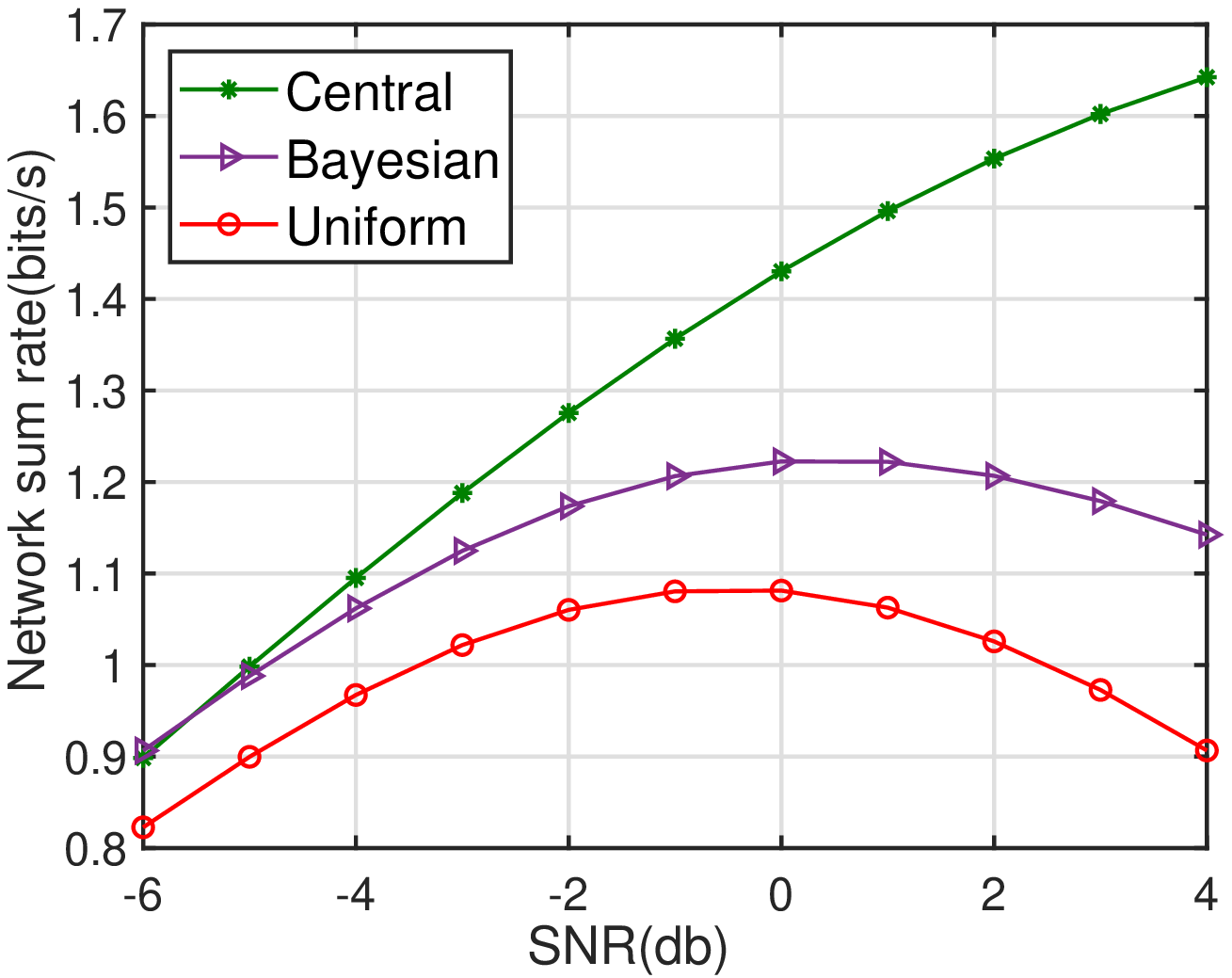}
\\{\footnotesize(a)}
\end{minipage}
\hfill
\begin{minipage}[t]{0.32\linewidth}
\centering
\includegraphics[width=\linewidth]{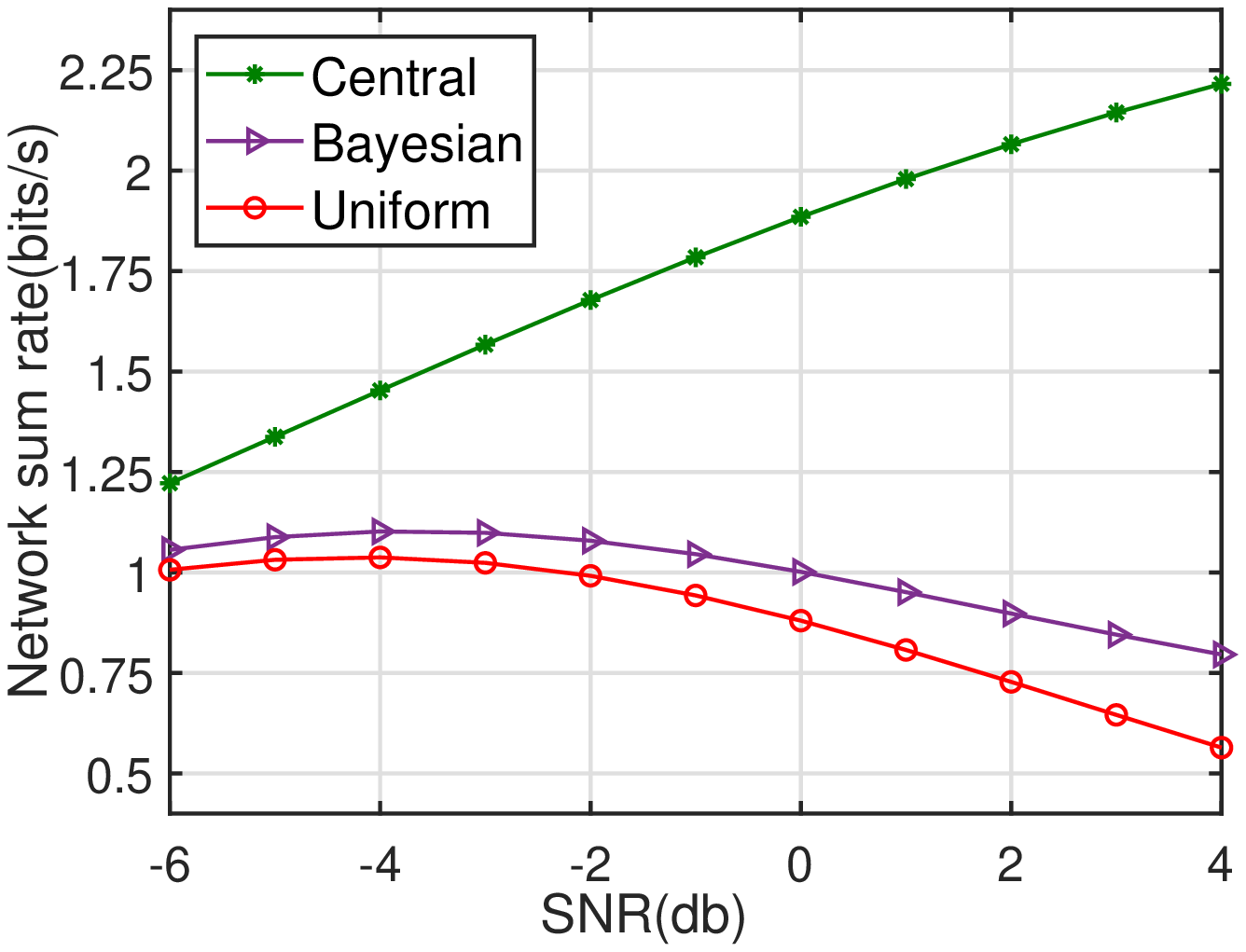}
{\footnotesize(b)}
\end{minipage}
\hfill
\begin{minipage}[t]{0.32\linewidth}
\centering
\includegraphics[width=\linewidth]{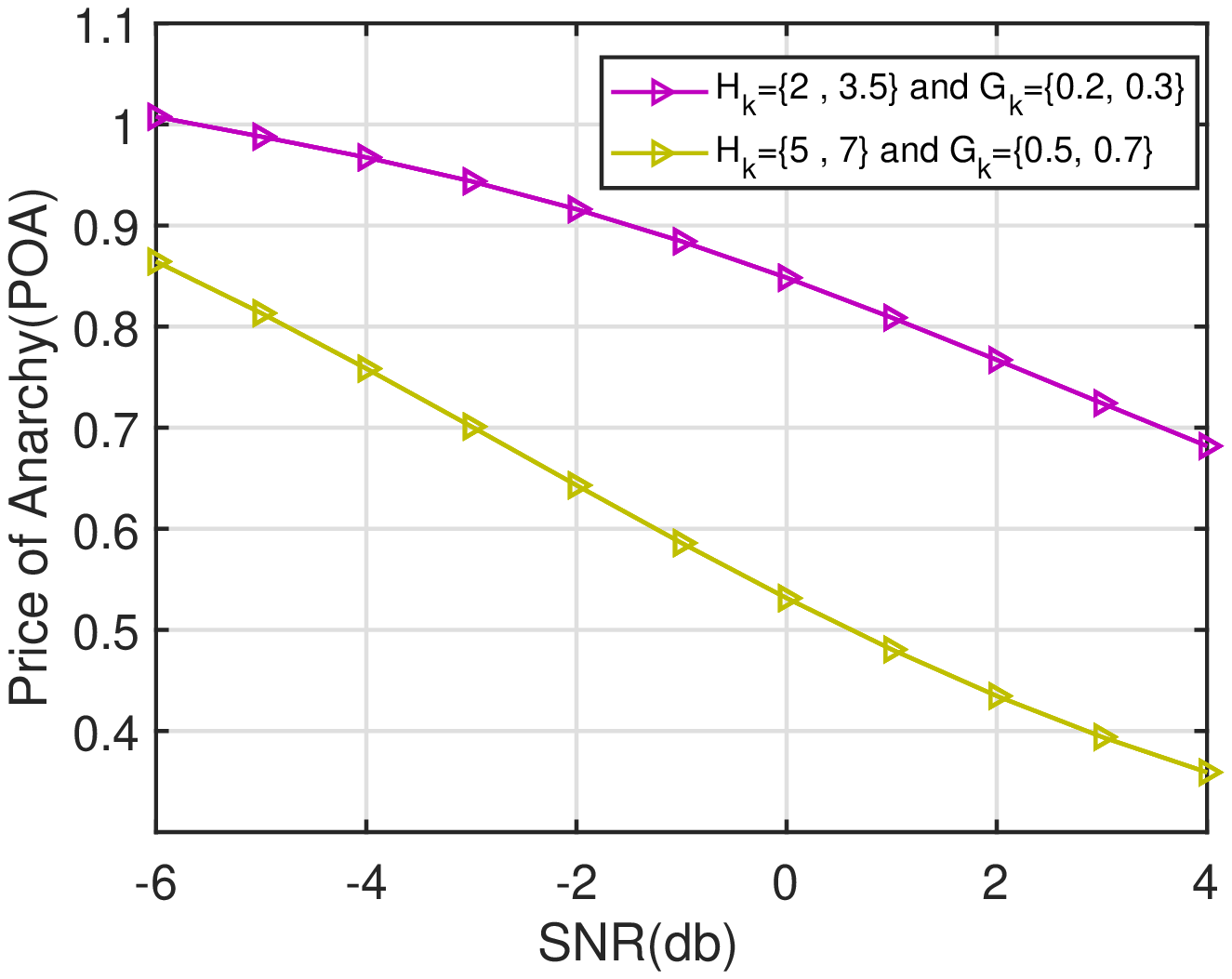}
{\footnotesize(c)}
\end{minipage}
\caption{(a) Sum rate vs $SNR$ for: (a) $H_{k}=\{2, 3.5\}$ and $G_{k}=\{0.2, 0.3\}$. (b) $H_{k}=\{5 , 7\}$ and $G_{k}=\{0.5,0.7\}$. (c) Performance of Bayesian Equilibrium}
\label{fig:efficiency}
\vspace{-0.6cm}
\end{figure*}
To assess the efficiency of the Bayesian equilibrium in terms of the network sum rate, in figure (\ref{fig:efficiency}), we plot the \emph{ergodic secrecy sum rate} and POA as a function of SNR. In figure (2a) we consider the same system parameters as specified above while in figure (2b) we have $H_k=\{5, 7\}$ and $G_k=\{0.5, 0.7\}$ while rest of the parameters remain unchanged. The ``Uniform" curve illustrates the sum rate when the user lacks channel state information, resulting in a uniform power allocation across all channel states by each mobile user. The ``Bayesian" curve represents the sum rate achieved at BE, where each user allocates power based on Algorithm (\ref{alg:loop}). Lastly, the ``Central" curve depicts the sum rate obtained when a centralized approach is employed to maximize the network sum rate, as explained in Section (IV-A). This curve serves as an upper bound for the BE. Finally, in figure (2c) we have plotted POA w.r.t to SNR for specified channel realizations. The objective is to achieve a POA value of one, indicating an efficient system at equilibrium. However, the figures clearly demonstrate the inefficiency of the system at equilibrium, which becomes more pronounced at higher SNR values.

Moreover, using a centralized approach for power allocation necessitates global CSI, which can be challenging to obtain, particularly in the presence of adversaries. Therefore, the significance of the Bayesian solution lies in its ability to develop secure self-organizing networks that reduce operational complexity and eliminate the requirement for global CSI.
\section{Conclusion}
We have addressed the problem of distributed power allocation in a multiple access wiretap channel, where global channel state information is limited. The problem is modeled as a Bayesian game, where each user is assumed to selfishly maximize his average \emph{secrecy capacity} with partial channel state information. We have proved that there exists a unique BE in the proposed game and developed a distributed iterative algorithm to characterize the BE. However, the network sum rate at BE exhibits inefficiency when compared to the centralized policy, especially in the high SNR regime. Hence, future direction involves exploring the implementation of pricing strategies or enabling strategic interactions among users to enhance the achievable \emph{secrecy rate} at BE.
\bibliographystyle{IEEEtran}
\bibliography{main}

\begin{thebibliography}{10}
\providecommand{\url}[1]{#1}
\csname url@samestyle\endcsname
\providecommand{\newblock}{\relax}
\providecommand{\bibinfo}[2]{#2}
\providecommand{\BIBentrySTDinterwordspacing}{\spaceskip=0pt\relax}
\providecommand{\BIBentryALTinterwordstretchfactor}{4}
\providecommand{\BIBentryALTinterwordspacing}{\spaceskip=\fontdimen2\font plus
\BIBentryALTinterwordstretchfactor\fontdimen3\font minus
  \fontdimen4\font\relax}
\providecommand{\BIBforeignlanguage}[2]{{%
\expandafter\ifx\csname l@#1\endcsname\relax
\typeout{** WARNING: IEEEtran.bst: No hyphenation pattern has been}%
\typeout{** loaded for the language `#1'. Using the pattern for}%
\typeout{** the default language instead.}%
\else
\language=\csname l@#1\endcsname
\fi
#2}}
\providecommand{\BIBdecl}{\relax}
\BIBdecl

\bibitem{ref_article1}
S.~Shamai and A.~D. Wyner, ``Information-theoretic considerations for
  symmetric, cellular, multiple-access fading channels.i,'' \emph{IEEE
  Transactions on Information Theory}, vol.~43, no.~6, p. 1877–1894, 1997.

\bibitem{ref_article2}
D.~N.~C. Tse and S.~V. Hanly, ``Multiaccess fading channels. i. poly-matroid
  structure, optimal resource allocation and throughput capacitie,'' \emph{IEEE
  Transactions on Information Theory}, vol.~44, no.~7, p. 796–2815, 1998.

\bibitem{ref_article3}
L.~Lai and H.~E. Gamal, ``The water-filling game in fading multiple access
  channels,'' \emph{IEEE Transactions on Information Theory}, vol.~54, no.~5,
  pp. 2110 -- 2122, 2008.

\bibitem{ref_article4}
G.~He, M.~Debbah, and E.~Altman, ``A bayesian game-theoretic approach for
  distributed resource allocation in fading multiple access channels,''
  \emph{EURASIP Journal on Wireless Communications and Networking}, 2010.

\bibitem{Nguyen2015Stackelberg}
N.~Duy~Duong, A.~S. Madhukumar, and D.~Niyato, ``Stackelberg bayesian game for
  power allocation in two-tier networks,'' \emph{IEEE Transactions on Vehicular
  Technology}, vol.~65, no.~4, pp. 2341 -- 2354, 2015.

\bibitem{tekin2007secrecy}
E.~Tekin and A.~Yener, ``Secrecy sum-rates for the multiple-access wire-tap
  channel with ergodic block fading,'' in \emph{45th Annual Allerton Conference
  on Communication, Control and Computing}, 2007, pp. 856--863.

\bibitem{shah2012achievable}
S.~M. Shah, V.~Kumar, and V.~Sharma, ``Achievable secrecy sum-rate in a fading
  mac-wt with power control and without csi of eavesdropper,'' in \emph{2012
  International Conference on Signal Processing and Communications
  (SPCOM)}.\hskip 1em plus 0.5em minus 0.4em\relax IEEE, 2012, pp. 1--5.

\bibitem{shah2016resource}
S.~M. Shah, A.~K. Chaitanya, and V.~Sharma, ``Resource allocation in fading
  multiple access wiretap channel via game theoretic learning,'' in \emph{2016
  Information Theory and Applications Workshop (ITA)}.\hskip 1em plus 0.5em
  minus 0.4em\relax IEEE, 2016, pp. 1--7.

\bibitem{Zhifan2021friendly}
Z.~Xu and M.~Baykal-Gürsoy, ``A friendly interference game in wireless secret
  communication networks,'' \emph{10th International Conference, NetGCooP},
  2021.

\bibitem{ref_article9}
J.~B. Rosen, ``Existence and uniqueness of equilibrium points for concave
  n-person games,'' \emph{Econometrica: Journal of the Econometric Society}, p.
  520–534, 1965.

\bibitem{tinh2022practical}
B.~T. Tinh, L.~D. Nguyen, H.~H. Kha, and T.~Q. Duong, ``Practical optimization
  and game theory for 6g ultra-dense networks: Overview and research
  challenges,'' \emph{IEEE Access}, 2022.

\bibitem{ref_article8}
S.~P. Boyd and L.~Vandenberghe, \emph{Convex optimization}.\hskip 1em plus
  0.5em minus 0.4em\relax Cambridge university press, 2004.

\bibitem{ref_article10}
J.~Papandriopoulos and J.~S.~Evans, ``Distributed algorithms for spectrum
  balancing in multi-user dsl networks,'' \emph{IEEE International Conference
  on Communications}, 2006.

\bibitem{ref_article11}
L.~An and P.~Tao, ``The dc (difference of convex functions) programming and dca
  revisited with dc models of real world nonconvex optimization problems,''
  \emph{Annals of Operations Research}, vol. 133, pp. 23--46, 2005.

\end{thebibliography}

\end{document}